\documentclass{desyproc}

\begin{document}
\title{Status Report of LPCC Forward Physics Group}

\author{{\slshape Oldrich Kepka}\\[1ex]
 	Institute of Physics, Academy of Sciences of the Czech Republic,\\
Na Slovance 2, 18221 Prague, Czech Republic}

\contribID{EDSBlois/2013/32}


\acronym{EDS'2013} 

\maketitle

\begin{abstract}
We report here about activities of newly formed working group aiming
to define a forward and diffractive physics program at the LHC for the
years to come.
\end{abstract}

\section{Introduction}
The physics program, for which the Large Hadron Collider (LHC) was build, is focused on the understanding of the electroweak symmetry breaking and search 
for possible Beyond Standard Model signal. The program is achieved with the use of detecting devices covering
mainly central regions in pseudo-rapidity ($|\eta|<5$), typically probing large transverse momentum of outgoing partons in proton-proton collisions (but also in proton-ion, ion-ion). The LHC is nominally configured to deliver highest instantaneous luminosities achivable giving a significant number of multiple proton-proton collisions per beam crossing, $\mu$.

\par
The nominal physics program can, however, be significantly extended if more information about the collision is gathered in the forward
region, beyond the reach of the central detectors of ALICE, ATLAS, CMS, and LHCb. There are several techniques to analyze properties of the high energy $pp$
collision. With sensitive devices installed hundreds of meters from the interaction point (IP) and just a few millimeters 
from the passing beam, it is possible to unambiguously identify scattered protons and measure their kinematics. Forward 
leading neutrons and photons can be measured by calorimeters installed  downstream from the IP, where charge particles 
are not observed, since they were deflected by the  LHC magnetic field upstream. Another possibility to probe the structure of hadronic activity is to 
cover the forward space with Forward Shower Counters to observe regions in pseudo-rapidity without any hadronic activity -- 
the so called rapidity gaps. 
\par 
The operation of detectors in forward region is, however, a complicated matter and requires a special attention. First, the impact of multiple 
proton-proton collisions (pile-up) is much more severe in the forward direction than at central pseudo-rapidities. Second, in order to optimize the acceptance and hence the physics reach for certain physics processes, it is desirable to tune the configuration of LHC magnets (different 
$\beta$* optics) and collect data in the so called special runs. Since the nominal LHC physics program requires the highest collected
luminosity possible, there is always a pressure to limit the number of special LHC runs, making requests of these runs uneasy. Any additional program therefore has to
be very well justified and planned. Third, since the active detectors are to approach LHC beams within the range of a few mm, technical solutions need to be developed, which imply no or small impact on the LHC machine and high luminosity running (beam induced background, unwanted beam dumps, etc).  Consequently, the approval of new projects or upgrades of the existing ones and their installation is more challenging. 
\par 
It was therefore realized that a common strategy between LHC experiments should be discussed in dedicated meetings over the course of one year, with a summary published in the CERN Yellow Report by mid-2014. The impact of the document should be large, the aim and strategy clearly outlined, and it should help to persuade the rest particle physics community to engage the forward physics program. The goal is to 
\begin{itemize}
   \item demonstrate the physics interest of forward physics and summarize the current results,
   \item define common strategy for running conditions optimal for forward physics (low-luminosity, special optics)
   \item provide clearer picture of the limitations \& overlaps of different experiments,
   \item engage theorists to help to bring ideas of new forward physics studies at the LHC.
\end{itemize}
The group was formed in May 2013 within the LHC Particle Physics Center~\cite{lpcc}.

\section{LPCC Forward Physics Working Group}
The LPCC Forward Physics Working Group is led by Nicolo Cartaglia (CMS, Torino) and Christophe Royon (ATLAS, Saclay), and is divided into four 
subgroups. The subgroups are categorized in such a way to best reflect the experimental challenges physics analyses 
will be facing.

\subsection{Low luminosity ($\sim 1\,\mathrm{pb}^{-1}$, $\mu<1$)}
This running scenario covers physics processes with large cross sections. It is expected that millions of single and double tagged events will be collected during these runs. They will allow a detailed investigation of diffractive cross section using both the proton tagging technique
or the rapidity gap method. For instance, the validity of the triple pomeron vertex model for a large range of diffractive masses, the applicability of multiple parton interaction model for diffractive events, universality of the fragmentation function in non-diffractive and diffractive events, effects of survival due to additional rescattering, etc.  In events with some hard perturbative scale, the partonic structure of the pomeron will be measured using for example single diffractive production of jets. 
The exclusive production (via multigluon exchange) of mesons ($\chi_c,\chi_b$, etc), or photoproduction ($J/\Psi/\Psi(2S), \Upsilon$) can also be measured, and provide access to very low-$x$ gluon in the proton and also proton survival factor. Events with forward neutrons detected in forward calorimeters add information about particle and energy flow in forward pseudo-rapidities,  and new measurements will improve models of hadron collisions in the Monte-Carlo generators. With the advancement of the tuning, it is hoped to obtain models which can be reliably extrapolated to
energies beyond the reach of the LHC, relevant for the modeling of cosmic ray showers. 

\subsection{ \bf Medium luminosity ($\sim 10-100\,\mathrm{pb}^{-1}$, $\mu\sim1$)}
Analyses falling into this category can exploit significantly more data, but on the other hand need to deal with small but non-negligible pile-up. The pile-up can be suppressed by the requirement of single interactions (rejecting events with more than one vertex reconstructed
from charged particles). Concerning the proton tagging, the pile-up will be suppressed employing the time-of-flight of the proton from the IP. This implies that the focus of this category are double tagged events, with protons seen on either side of the IP reconstructed with moderate timing resolution. Background coming from the combinatorial background due to overlap of multiple single diffractive events does not have a timing information compatible with the vertex position reconstructed in the central tracker. The aim is to measure the structure of the pomeron in detail (including flavor), but also to search for  non-DGLAP evolution of the pomeron parton density functions. It is possible that some exclusive channels might be observed (such as exclusive production of pion pairs) with proton tags.

\subsection{ \bf High luminosity ($>10\,\mathrm{fb}^{-1}$, nominal running conditions)}
In contrast with the preceding scenarios, analyses considered in this category do not require any sort of special running (neither in terms of optics nor the level of pile-up). It is intended that forward detectors will be operational during the nominal data taking of LHC experiments. It is therefore crucial to utilize various techniques to suppress the level of pile-up. The timing information outlined above is the key ingredient in these measurements. However, it was found that the timing itself does not provide a sufficient background rejection. Therefore, the programme consists of several exclusive measurements in which kinematic correlation between objects in the central detector and protons in the forward detectors helps to suppress the background. Note that since in double pomeron exchange, some energy is lost in pomeron remnants, the process is not exclusive, and requires data to be collected with lower pile-up. One of the interesting physics applications of the forward proton tagging is the precise determination of Quartic Gauge Boson couplings. It has been shown that the $\gamma\gamma\rightarrow WW/ZZ/\gamma\gamma$ (coherent radiation of photons off protons) can be constrained with higher precision than using the conventional methods. Also, it has been shown that the proton tagging can be used to constrain the exclusive production (due to exchange of gluons) of jets and hence provide a substantial reduction of uncertainties associated with the exclusive production of the Higgs boson.

\subsection{ \bf Technical working group}

The purpose of the technical working group is to discuss technical aspects regarding the development of proton tagging and integration inside the LHC which are common between different experiments around the LHC ring.

\section{Conclusion}

The aim of this report was to inform the community about the newly formed LPCC Forward Physics Working Group. The group is preparing a CERN Yellow report to summarize and plan diffractive and exclusive measurements which should be performed after the first long shutdown of LHC. It is possible to sign up for {\it lhc-fwdlhcwg} mailing list at~\cite{egroup}.

\section{Acknowledgments}
This work is partially supported by WP8 of the hadron physics program of
the 8th EU program period.

%


\begin{footnotesize}

\end{footnotesize}
\end{document}